\documentclass[epjCONF, onecolumn]{svjour}
\usepackage{graphics}
\usepackage[varg]{txfonts} 
\usepackage[latin1]{inputenc}
\session-title{Assembling the Puzzle of the Milky Way}
\begin{document}
\title{Orbits of LMC/SMC with recent ground-based proper motions}
\author{Kate\v{r}ina Barto\v{s}kov\'{a}\inst{1,2}\fnmsep\thanks{\email{bartoskova@physics.muni.cz}} \and Bruno Jungwiert\inst{2,3} \and Adam R\r{u}\v{z}i\v{c}ka\inst{2,4} \and Edgardo Costa\inst{5}}
\institute{Department of Theoretical Physics and Astrophysics, Masaryk University, Brno, Czech Republic; \and Astronomical Institute, Academy of Sciences of the Czech Republic, Prague, Czech Republic; \and Faculty of Mathematics and Physics, Charles University in Prague, Czech Republic; \and Institut f\"{u}r Astronomie der Universitat Wien, Wien, Austria; \and Departamento de Astronomia, Universidad de Chile, Chile}
\abstract{
In recent years, with new ground-based and HST measurements of proper motions of the Magellanic Clouds being published, a need of a reanalysis of possible orbital history has arisen. As complementary to other studies, we present a partial examination of the parameter space  -- aimed at exploring the uncertainties in the proper motions of both Clouds, taking into account the updated values of Galactic constants and Solar motion, which kinematically and dynamically influence the orbits of the satellites. In the chosen setup of the study, none of the binding scenarios of this pair could be neglected.
} 
\maketitle
\section{Regarding the variety of Proper Motions}
\label{intro}
Proper motions (PM) from HST measurements \cite{Ref1,Ref2,Ref3} increased possible velocities of the Magellanic Clouds with the respect to the Galactic center, closer to or exceeding the escape velocity. Subsequent orbital studies came with a theory of the first orbital passage \cite{Ref7} and challenged the plausibility of a traditionally accepted bound satellite history.
Recently, using the updated position of the Sun and $V_{\mathrm{LSR}}$ \cite{Ref12,Ref13}, several works show, that even with the HST proper motions, the Magellanic Clouds are more likely to be bound to the MW \cite{Ref9,Ref10,Ref11}. New ground-based proper motion measurements \cite{Ref4,Ref5,Ref6}, which use different approach to the PM estimation, bring back the possibility of having even lower velocities.

In this paper, we present a simple illustrative study of the past orbital history, taking into account the mentioned updated rotation rate and the recently published velocity vector of the Sun \cite{Ref8}. A set of orbits of two softened point masses approximating the LMC and SMC in the rigid analytical potential of the Milky Way (see Table \ref{tab:1}), with dynamical friction included, has been calculated. 
 \begin{figure}\sidecaption
\resizebox{0.63\columnwidth}{!}{%
\includegraphics{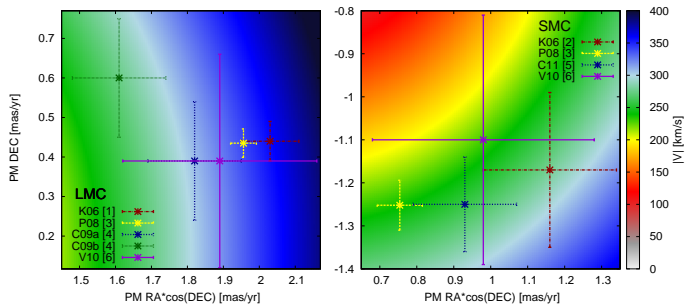}
}
\caption{Range of Centre-of-mass proper motions from literature \cite{Ref1,Ref2,Ref3,Ref4,Ref5,Ref6}, and derived $|\vec{V}|$, which has overall lower values assuming updated Galactic constants (here: $R_{\odot}~=~8.4$~kpc, $V_{\mathrm{LSR}} = 242$~km/s, $\vec{V}_{\odot} = (11.10, 12.24, 7.25)$~km/s) in comparison with the standard IAU value of $V_{\mathrm{LSR}}=220$ km/s, and with the obsolete Hipparchos Solar motion.\smallskip}
\label{fig:1}       
\end{figure} Simulations within the presented set differ only by the initial velocity vector corresponding to the sweep over the uncertainty of the PM space (see Fig. \ref{fig:1}). Some graphical outputs of the results are depicted in Fig. \ref{fig:2}.
\label{sec:1}

\begin{table}
\label{tab:1}       
\begin{tabular}{c  c  | c  c  c | c c c}
  $r_{\mathrm{bulge}}$ & $M_{\mathrm{bulge}}$ & $R_{\mathrm{disk}}$ & $z_{\mathrm{disk}}$  &  $M_{\mathrm{disk}}$ & $c$ & $M_{\mathrm{vir}}$ & $r_{\mathrm{vir}}$ \\
\hline 
 $0.7$ kpc & $1.5 \cdot 10^{10}$ M$_{\odot}$ & $4.0$ kpc & $0.26$ kpc  & $5.0 \cdot 10^{10}$ M$_{\odot}$  & $12$ & $2.0 \cdot 10^{12}$ M$_{\odot}$ & $325$ kpc\\
\end{tabular}\smallskip
\caption{The Milky Way model: Hernquist bulge, Miyamoto-Nagai disk and NFW halo. Virial mass and radius are set to roughly correlate with the updated $V_{\mathrm{LSR}}$, similarly to \cite{Ref9} approach.}
\end{table}

\begin{figure}
\resizebox{1.0\columnwidth}{!}{%
\includegraphics{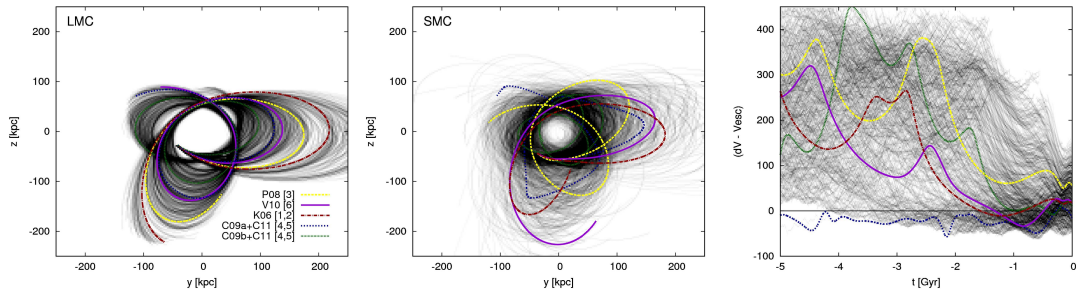} 
}
\caption{From the left, \textbf{a)} and \textbf{b)}: Distribution of LMC and SMC orbits during the backward integration (to $-5$~Gyr) projected onto the $y-z$ plane. Selected solutions for the mean values of measured proper motions are plotted in colors, according to Fig. \ref{fig:1}; \textbf{c)} Right panel: Evolution of the difference between the amplitude of the relative velocity LMC--SMC and the escape velocity from the LMC potential -- for this simple approach -- the Magellanic clouds being softened points with classical assumptions of masses: $(M_{\mathrm{LMC}}, M_{\mathrm{SMC}}) = (2.0, 0.3)\cdot 10^{10} \mathrm{M}_{\odot}$. Statistically significant subset of results occupies the space with velocities lower then $V_{\mathrm{esc}}$, still around the time of $-1$~Gyr. Thus SMC may form the bound binary with LMC in similar configurations. Some of the results show long-term binding relationship -- a representative one may correspond to the blue mean value orbit \cite{Ref4,Ref5}.
}
\label{fig:2}       
\end{figure} \noindent Evidently, such results indicate Magellanic Clouds being bound to the Milky Way. The other question is, whether they can form a bound binary by themselves (possible similar formation history), or the SMC is an independent satellite, thus unbound all the time or later captured by LMC during close encounters (as proposed eg. by \cite{Ref11}). In the chosen setup within the wide range of PMs, all basic scenarios exist (see Fig. \ref{fig:2}). However, the decision on the long-term binding relationship is sensitive to the used range of PM values and to the treatment of gravity between the Clouds -- and therefore should be explored in a more realistic study.

\begin{acknowledgement}
 This project is supported by grants No. 205/08/H005 (Czech Science Foundation), MUNI/A/0968/2009 (Masaryk University in Brno), research plan AV0Z10030501 (Academy of Sciences of the Czech Republic), LC06014 (Center for Theoretical Astrophysics, Czech Ministry of Education), GACR P209/11/P699 (The Czech Science Foundation), EC acknowledges support by FONDECYT (Proy. No. 1050718 and 1110100).
\end{acknowledgement}

\end{document}